# Angular structure factor of the hexatic-B liquid crystals: bridging theory and experiment

Ivan A. Zaluzhnyy,[a,b] Ruslan P. Kurta,[c] Michael Sprung,[a] Ivan A. Vartanyants,[a,d†] Boris I.Ostrovskii[e,f‡]

We report results from X-ray scattering studies of the angular structure factor of liquid crystal hexatic-B films. According to the sixfold rotational symmetry of the hexatic-B phase, its characteristic scattering splits into six reflections. The shape of the radial and angular cross-sections of these reflections and their temperature evolution are analyzed. We find that over a wide temperature range of the hexatic-B phase existence the angular profiles of the in-plane X-ray scattering are well fitted by the Voigt function, which is a convolution of the Gaussian and Lorentzian functions. This result is supported by the known theoretical considerations of the hexatic structure factor below the smectic-hexatic phase transition temperture. Similar predictions for the angular shape of the hexatic peak in the vicinity of the smectic-hexatic phase transition temperature follow from the multicritical scaling theory of the hexatic-B phase in three dimensions. We find that the specific shape of the hexatic structure factor can be explained by the interplay of two distinct contributions to the free energy of the system, a liquid-like density term and a coupling term between the bond-orientational order and short-range density fluctuations.

## 1 Introduction

The hexatic-B (Hex-B) liquid crystal (LC) is a unique stacked three-dimensional (3D) analogue of the notable two-dimensional (2D) hexatic phase that was predicted as an intermediate state in the process of 2D crystal melting.[1-3] The Hex-B phase is closely related to the well-known smectic-A (Sm-A) phase, therefore LC substances often exhibit a temperature induced Sm-A – Hex-B phase transition. Both the Hex-B and Sm-A phases consist of a stack of parallel equidistant molecular layers, in which elongated molecules are oriented on average along the layer normal, and exhibit the short-range positional correlations within the layers (Fig. 1(a,b)). The Hex-B phase differs from the Sm-A phase by the presence of a long-range bond-orientational (BO) order.[4-6] The Hex-B phase is a truly 3D system: the orientation of intermolecular bonds is the same in all layers in the stack, while the shear modulus in the plane of layers is zero, so there are no positional correlations between them (Fig. 1 (b)).[2-5] The BO order in the Hex-B phase is characterized by a sixfold rotational symmetry and can be described by the complex-valued local ordering field $\Psi(r) = |\Psi(r)|e^{i6\psi(r)}$, where $|\Psi(r)|$ is a magnitude of the two-component BO order parameter and $\psi(r)$ is its phase that corresponds to an angle between the intermolecular bonds and some reference axis. In addition to classical thermotropic LCs, the 3D hexatic phase can be also found in some other systems, such as lipids,[7-8] DNA solutions,[9-10] sanidic mesophases,[11] self-assembled micellar polymers[12] and columnar assemblies of colloidal discs.[13]

The structure of a homogeneous system, such as Sm-A or Hex-B phase, can be described in terms of the pair density correlation function $G(r) = \langle \delta\rho(r + R)\delta\rho(R)\rangle_R$, which gives the probability to find a particle at the separation $r$ from an arbitrary chosen reference particle positioned at $R$.[14,15] Here $\delta\rho(r)$ is a local deviation of the electron density from the average value, and the averaging $\langle...\rangle_R$ is performed over the positions of a reference particle $R$. The structure factor $S(q)$ measured in an X-ray scattering experiment is a Fourier transform of the pair density correlation function, and therefore it can be written as

$$S(q) = \int G(r)e^{iqr}dr = \langle|\delta\rho(q)|^2\rangle, \quad (1)$$

where $q$ is the momentum transfer vector and $\delta\rho(q)$ is a Fourier transform of $\delta\rho(r)$. Hence, the information on the character of the orientational and positional order, as well as the symmetry of the system can be obtained either directly from $S(q)$ or by reconstructing the pair density correlation function $G(r)$ from the scattering data.[16,17]

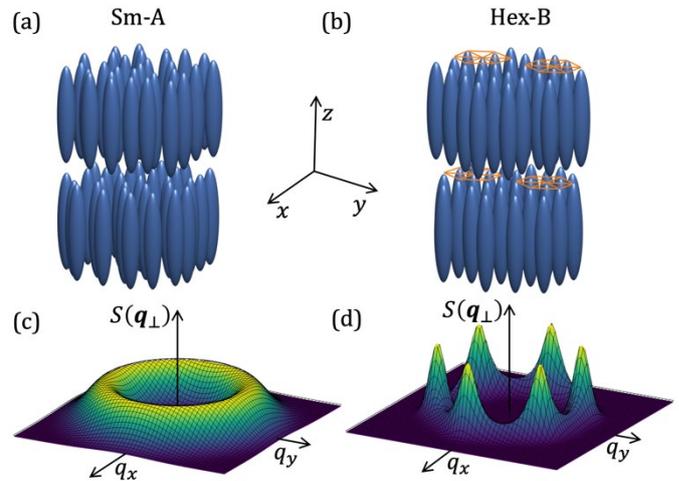

**Fig. 1.** (a-b) Layered molecular structure of the Sm-A and Hex-B phases. Two molecular layers are shown with the liquid-like short-range positional order within a layer. Orange hexagons indicate the bond-orientational order in the Hex-B phase. (c-d) Corresponding structure factor $S(q_\perp)$ for the Sm-A (c) and Hex-B (d) phases.

The general feature of all layered (smectic, hexatic) LC phases in reciprocal space is a set of sharp $(00n)$ quasi-Bragg peaks in the direction along the layer normal $(q_z)$ at wave-vector transfer $q_{zn} = 2\pi n/d$ (where $d$ is a layer spacing). The in-plane positional order is short-range and can be determined from the position and width of the broad diffuse peaks in the $q_\perp$ plane $(q_\perp = (q_\perp, \varphi)$ in polar coordinates) (Fig. 1(c,d)). In the following we consider the structure factor $S(q_\perp)$ in the plane of the layers ($q_z \ll 2\pi/d$), which allows one to identify the Hex-B phase and to analyse its structural features.

In our previous studies of the hexatic films, we have obtained numerous data on the structure factor $S(q_\perp)$ over the range of existence of the Hex-B phase, including the vicinity of the phase transition from a high temperature Sm-A to the Hex-B phase.[18-20] In this work, we go beyond the standard analysis of the in-plane structure factor of the Hex-B phase and present the X-ray data from thick free-standing films of four different LC materials. We confirm that the radial profiles of the scattering peaks are well approximated by the Lorentzian functions as expected for systems with a liquid-like in-plane order.[16-17] The angular profiles of the scattering peaks in the Hex-B phase are more complex and can be described by a combination of the central Gaussian peak with the extended Lorentzian wings. Our experimental data show that the angular profiles of the hexatic reflections are well fitted by the Voigt profile, which is a convolution of the Gaussian and Lorentzian functions, at all temperatures within the region of the Hex-B phase existence. The relative contribution to the angular scattering profile from both functions depends on the temperature deviation $\Delta T = T - T_C$ from the Sm-A to Hex-B transition point $T_C$ and, accordingly, on the strength of the BO order.

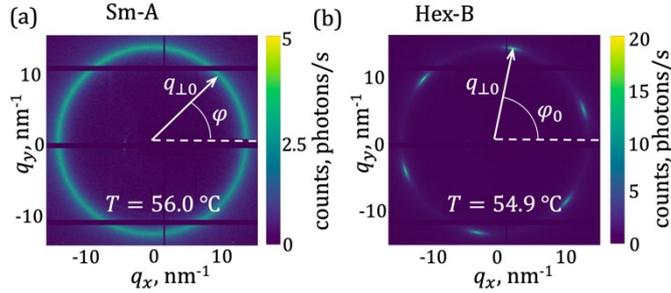

**Fig. 2.** Experimentally measured X-ray scattered intensity $I(q_\perp) = I(q, \varphi)$ for the LC compound 54COOBC at temperature T=56 °C in the Sm-A phase (a) and T=54.9 °C in the Hex-B phase (b). $q_{\perp 0}$ is the radius of the diffuse scattering ring, $\varphi$ is an angular variable, $\varphi_0$ defines the angular position of the diffraction peaks in the Hex-B phase.

The obtained result is of a fundamental origin and reflects the intrinsic symmetry of the Hex-B structure factor. It agrees very well with the theoretical study of Aeppli and Bruinsma[21] where they have shown that the angular cross-section of the hexatic structure factor in a 3D hexatic LC can be described by a convolution of the Gaussian and Lorentzian functions (see also[22]). Their model is based on the interaction between the BO order parameter $\Psi(r)$ and the short-range density fluctuations $\delta\rho(r)$, introduced by a cross-term of the Landau functional depending on both fields $\Psi(r)$ and $\delta\rho(r)$. Similar predictions for the angular shape of the structure factor follow from the multicritical scaling theory (MCST) developed by Aharony et al. for the 3D Hex-B phase.[23] Despite the fact that these theoretical predictions were known for several decades, they have never been thoroughly verified experimentally. To the best of our knowledge the first experimental work in this direction[24] was limited by only two LC compounds and the comparison with the theories of Aeppli and Bruinsma[21] and MCST[23] was lacking. In this work, we analyse in detail the angular shape of the diffraction peaks in the Hex-B phase and compare the experimental data with the existing theoretical models.

## 2 Bond-orientational order and the hexatic structure factor

The in-plane correlations between the molecules in the Sm-A phase give rise to a broad diffuse scattering ring in the $q_\perp$ plane, Fig. 2(a). Due to the short-range positional order between the molecules in a smectic layer, the radial cross section of the ring has a Lorentzian profile with a half width at half maximum (HWHM) $\gamma$. This allows one to define a characteristic length $\xi = 1/\gamma$ over which positional correlations between the molecules in a layer decay exponentially. The radius of the ring is $q_{\perp 0} \cong 4\pi/a\sqrt{3}$, where $a$ is the average lateral molecular separation. The emergence of the BO order in the Hex-B phase breaks the angular isotropy of the in-plane structure factor and leads to a six-fold modulation of the in-plane scattering, Fig. 2(b). A Fourier expansion of the azimuthal scattering profile into cosine series gives [5]

$$I(q_{\perp 0}, \varphi) = I_0 \left[1 + 2 \sum_{m=1}^{+\infty} C_{6m} \cos(6m(\varphi - \varphi_0))\right], \quad (2)$$

where $I_0$ is the angular-averaged scattered intensity, $\varphi$ is the angle measured along the arc of the diffraction pattern, and $\varphi_0$ corresponds to the orientation of the diffraction pattern relative to the reference axis in the $q_\perp$-plane. The coefficients $C_{6m}$ are usually considered as the BO order parameters measuring the degree of the BO ordering in the Hex-B phase.[5,18] In the Sm-A phase, the diffuse ring is uniform leading to $C_{6m} = 0$ (Fig. 2(a)).

Upon cooling, just below the Sm-A – Hex-B phase transition temperature, only a subtle modulation of the intensity along the scattering ring appears. On further cooling, the scattering ring splits into six individual reflections when the BO order becomes strong enough (Fig. 2(b)). This corresponds to development of the BO order, which manifests itself both in an increase of the number of the nonzero BO order harmonics $C_{6m}$, and their magnitude.[5,18-20] These X-ray reflections are asymmetric, each spot is narrow in the radial direction (but still diffuse) and elongated in the azimuthal direction.

Simultaneously with these changes in the azimuthal distribution of the scattered X-ray intensity, the radial width of the scattering peaks also decreases. This indicates a coupling between the positional correlations $G(r)$ defined by the correlation length $\xi$, and the BO order.[19,20] The azimuthal scattering profile $I(q_{\perp 0}, \varphi)$ in eqn (2), in the limit of infinite $m$ when all coefficients $C_{6m}$ increase to unity transforms to six sharp peaks in the form of delta-functions, $I \propto \sum_{n=1}^{6} \delta(\varphi - \varphi_0 - \pi n/3)$. This means that at a certain stage the increase of the positional in-plane order might lead to the formation of the 2D lattices that are locked together into the 3D crystalline structures. This should correspond to the formation of a hexagonal crystal which is characterized by a symmetrical sixfold pattern of true Bragg peaks.

As we mentioned above, the X-ray scattering technique measures the Fourier-transformed density– density correlation function $G(r)$ (see eqn (1)). The appearance of the BO order in the system by itself does not change the local density fluctuations responsible for X-ray scattering. The measurements of the evolution of the BO order are possible only due to coupling between the liquid density fluctuations $\delta\rho(r)$ and the hexatic order parameter $\Psi(r)$.[21,22] This situation is similar to nematic LCs, where the long-range orientational order of the long axis of the molecules couples with the liquid-like density.[16] This leads to the following generic expression for the hexatic structure factor[21,22,25,26]

$$S(q_\perp) = \langle|\delta\rho(q_\perp)|^2\rangle \propto \langle \frac{k_B T}{A(q_\perp) + F(q_\perp, |\Psi|, \cos[6(\varphi - \varphi_0)])} \rangle. \quad (3)$$

Here $k_B T$ is a thermal energy with $k_B$ being a Boltzmann constant, and $A(q_\perp) = \xi^{-2} + (q_\perp - q_{\perp 0})^2$ is the contribution of the short-range positional order into the free energy, where $q_{\perp 0}$ is the preferred in-plane wave vector. The term $A(q_\perp)$ has a minimum at $q_\perp = q_{\perp 0}$, which corresponds to the maximum of the structure factor $S(q_\perp)$. The second term in the denominator of eqn (3), $F(q_\perp, |\Psi|, \cos[6(\varphi - \varphi_0)])$, depends on the BO order and takes into account the coupling between the BO order and the positional order in the hexatic phase. The average on the right-hand side of eqn (3) is over different configurations of the BO order. In accordance with the symmetry of the Hex-B phase, eqn (3) implies six maxima in the structure factor $S(q_\perp)$ at the reciprocal ring $|q_\perp| = q_{\perp 0}$. The expression for $A(q_\perp)$ suggests the Lorentzian profile of the scattering peak in the radial direction with a HWHM $\gamma$ inversely proportional to the positional correlation length ($\gamma = 1/\xi$), which is typical for the LC phases with a short range in-plane positional order.[16,17] The exact functional dependence of the term $F(q_\perp, |\Psi|, \cos[6(\varphi - \varphi_0)])$ is unknown, however some predictions regarding the shape of the angular (azimuthal) component of the Hex-B structure factor still can be made.

Thus, Aeppli and Bruinsma in their work[21,22] derived the structure factor of the Hex-B phase using the Landau free energy expansion that combines a liquid-like density term and a term describing the coupling between the BO order and the short-range density fluctuations (see Supplementary Information, section 1). Using reasonable assumptions, the authors analysed eqn (3) and have shown that the Hex-B structure factor crucially depends on the mean square amplitude of fluctuations of the phase of the BO order parameter, $\langle\delta\psi(r)^2\rangle$. It was shown that well below the Sm-A – Hex-B phase transition point, where the magnitude of $|\Psi(r)|$ can be considered constant and the fluctuations of the phase $\langle\delta\psi^2\rangle$ are small, the angular cross-section of the hexatic structure factor can be written as

$$S(q_{\perp 0}, \varphi) \propto \int_{-\infty}^{+\infty} d\theta \frac{e^{-\frac{(\theta-\varphi)^2}{2\langle\delta\psi^2\rangle}}}{\kappa^2 + \theta^2}, \quad (4)$$

where the value of the parameter $\kappa$ is defined by the short-range positional correlations within a molecular layer and the coupling between the positional order and the BO order.

Eqn (4) states that the azimuthal cross section of the hexatic peak can be described by a Voigt function

$$V(\varphi; \sigma, \kappa) = G(\varphi; \sigma) * L(\varphi; \kappa), \quad (5)$$

that is a convolution (denoted by asterisk) of a Gaussian function

$$G(\varphi; \sigma) = \frac{1}{\sqrt{2\pi\sigma^2}} e^{-\frac{\varphi^2}{2\sigma^2}} \quad (6)$$

with $\sigma^2 = \langle\delta\psi^2\rangle$, and a Lorentzian function

$$L(\varphi; \kappa) = \frac{\kappa}{\pi(\varphi^2 + \kappa^2)}. \quad (7)$$

In contrast to many other applications, including X-ray crystallography, where the Voigt function appears as a result of finite resolution of the experimental instruments, here it originates from the specific coupling between the BO order and the short-range density fluctuations in the hexatic phase. The Gaussian contribution arises from the fluctuations of the BO order and is determined by the parameter $\langle\delta\psi^2\rangle$, while the Lorentzian counterpart appears due to the coupling between the positional and BO orders and is determined by the value of the parameter $\kappa$ (see Supplementary Information, section 1).

Aeppli and Bruinsma[21] also analyzed the radial cross section of a hexatic structure factor and showed that far away from the Sm-A – Hex-B phase transition region the radial line shape can be well described by the Lorentzian profile with a HWHM $\gamma = 1/\xi$. However due to strong fluctuations of the phase of the BO order parameter in the vicinity of the Sm-A – Hex-B phase transition point, the radial intensity profile changes its shape and is better approximated by the square root of a Lorentzian.[21,27]

Another way of theoretical description of the BO order in hexatic LCs is the multicritical scaling theory (MCST) developed by Aharony et al,[23] and based on the proximity of the Sm-A – Hex-B phase transition to the tricritical point. The authors have shown that the BO parameters $C_{6m}$ defined in eqn (2) satisfy the following scaling law

$$C_{6m} = (C_6)^{\sigma_m}, \quad (8)$$

where

$$\sigma_m = m + \lambda m(m-1). \quad (9)$$

Theoretical considerations[23,28,29] suggest that the coefficient $\lambda$ is equal to $\lambda = 1$ for 2D hexatic films, and $\lambda = 0.3$ for the 3D systems. These predictions are in good agreement with the experimental results.[5,30-33] In our recent publications we have applied the angular X-ray cross-correlation analysis (AXCCA)[34-36] to study the BO order in the single-domain Hex-B films. By using this approach, the individual harmonics $C_{6m}$ of the BO order in the Hex-B phase can be directly determined without applying a fitting procedure to the angular intensity distributions. Our results support the validity of the MCST for LCs of various compositions.[19,20]

It can be shown that the MCST contains in a hidden form the information about the shape of the angular structure factor in the Hex-B phase. Thus, for 2D hexatics, $\lambda = 1$ and $\sigma_m = m^2$ in eqn (9), which allows to rewrite the scaling law (8) as $C_{6m} = e^{-18m^2\sigma^2}$, where the parameter $18\sigma^2 = -\ln C_6$. This expression for the Fourier coefficients suggests the Gaussian shape of the scattering peaks in the direction along the ark[29,30,37] (see Supplementary Information, section 2).

In the 3D hexatic phase, $\lambda \approx 0.3$ that corresponds to a more complex profile of the structure factor in the azimuthal direction. In this case, the scaling law of MCST can be rewritten as a product

$$C_{6m} = (C_6)^{\sigma_m} = e^{-18m^2\sigma^2} e^{-6|m|\kappa}, \quad (10)$$

where the parameters $\sigma$ and $\kappa$ depend on the value of the coefficient $C_6$ in the angular Fourier series and the parameter $\lambda$ (see Supplementary Information). The first term $\exp(-18m^2\sigma^2)$ in eqn (10) corresponds to the Fourier transform of a Gaussian function, and the second term $\exp(-6|m|\kappa)$ corresponds to the Fourier transform of a Lorentzian function. In this case, the convolution theorem states that the angular profile of a single hexatic peak is a Voigt function.

## 3 Experimental results and discussion

The X-ray scattering experiments were carried out at the coherence beamline P10 of the PETRA III synchrotron source at DESY. The details of the X-ray setup, LC materials, sample preparation, and experimental conditions can be found in refs.[18-20,38] Upon cooling, all LC materials undergo the Sm-A – Hex-B phase transition. We have studied thick (with the thickness of $1 - 15$ μm) free standing films of four different LC compounds: 75OBC[18,20,24], 3(10)OBC[19,20,24], PIRO6[20] and 54COOBC[38].

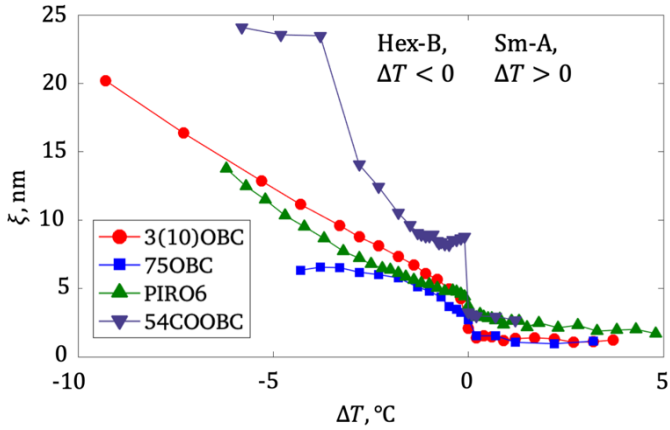

**Fig. 3.** Temperature dependence of the positional correlation length $\xi$. The temperature $\Delta T = T - T_C$ is measured relative to the phase transition point.

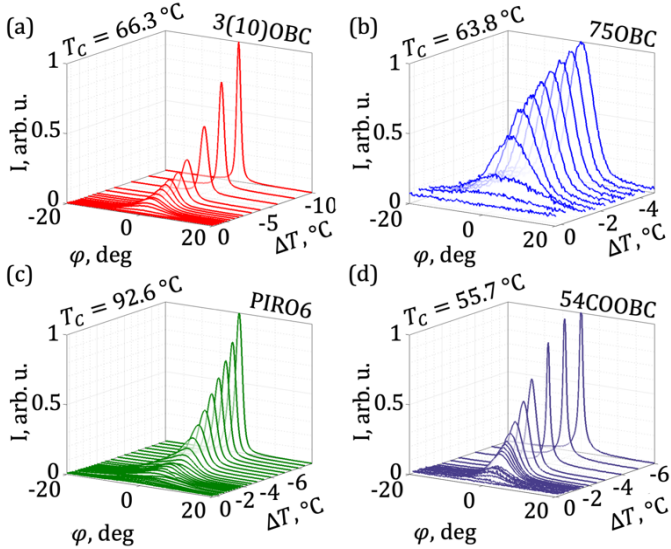

**Fig. 4.** Azimuthal dependence of the normalized scattered intensity $I(\varphi)$ of a single hexatic peak for four different LC compounds in the vicinity of the Sm-A – Hex-B phase transition and in the Hex-B phase: (a) 3(10)OBC, (b) 75OBC, (c) PIRO6, and (d) 54COOBC. The phase transition temperature $T_C$ is indicated for each compound.

The free-standing films are especially suitable for an X-ray scattering study: they have a controlled thickness, the films are not influenced by substrate interactions and their free surfaces promote an almost perfect alignment of the smectic layers.[39,40] The examples of the experimentally measured scattered X-ray intensity for 54COOBC compound are shown in Fig. 2. The azimuthally uniform diffuse ring in Fig. 2(a) corresponds to the liquid-like structure of the smectic layers in Sm-A phase. Upon cooling to the Hex-B phase, the scattering ring splits into six distinct arcs manifesting the appearance of the BO order within the large-scale single hexatic domains (Fig. 2(b)). The observation of the 2D diffraction pattern with six symmetrical spots indicates that in all layers in the stack, the bond angles between the neighbouring molecules are the same (Fig. 1(b)). This means that the average phase of the BO order parameter $\langle \psi(r) \rangle$ has the same value for all layers in the film. In our X-ray experiments, we have measured the intensity distribution within the single hexatic reflections in the radial and azimuthal directions. The resolution in our experiments was determined by the pixel size of the area detector for the given sample-detector distance. It was high enough ($\Delta q_\perp^{res} < 0.04$ nm$^{-1}$, $\Delta \varphi^{res} < 6 \cdot 10^{-4}$ deg) to resolve the profiles of the narrowest X-ray scattering lines both in the radial and azimuthal directions.[19,20] Below, we, first, shortly discuss the radial profile of the Hex-B structure factor, and then focus on its angular shape.

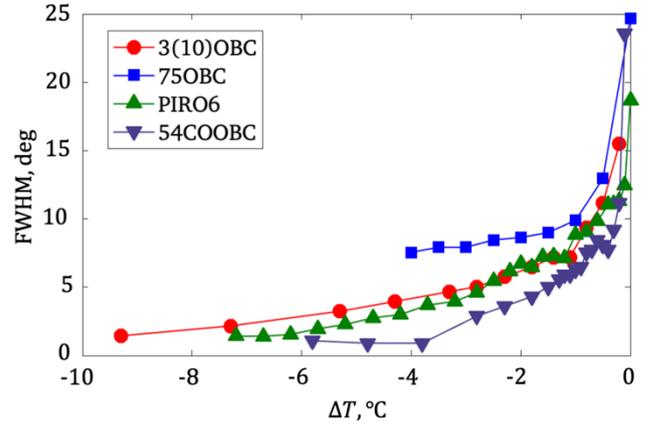

**Fig. 5.** Temperature dependence of the azimuthal width (FWHM) of a single hexatic peak for different LC compounds.

### 3.1 In-plane positional order and the radial intensity distribution

The in-plane positional order in the Sm-A and Hex-B phases of various LC compounds was studied using the shape analysis of the radial cross-sections of the diffraction patterns. The intensity distribution $I(q_\perp, \varphi_0) = I(q_\perp)$ in the radial direction through the maximum of one of the sixfold diffraction peaks in the Hex-B phase was averaged over different positions in the film and then fitted with the Lorentzian and square root Lorentzian (SRL) functions. In the case of 3(10)OBC and PIRO6 samples the SRL function gives better approximation for $I(q_\perp)$ in the temperature interval $-0.8$ °C $\leq \Delta T \leq 0$ °C and $-2.7$ °C $\leq \Delta T \leq 0.3$ °C, respectively, which is in the close vicinity of the Sm-A–Hex-B phase transition temperature. Outside of this range the Lorentzian function fits the experimental data better.[20] For the 75OBC sample, the quality of the data does not allow us to unambiguously determine which function, the Lorentzian or SRL, provides a better fit. In the case of 54COOBC films the situation is even more ambiguous, due to the presence of the two-phase coexistence region near the first-order Sm-A – Hex-B phase transition in this compound.[38] For consistency, below we present the results of the fitting with the Lorentzian function for all temperatures.

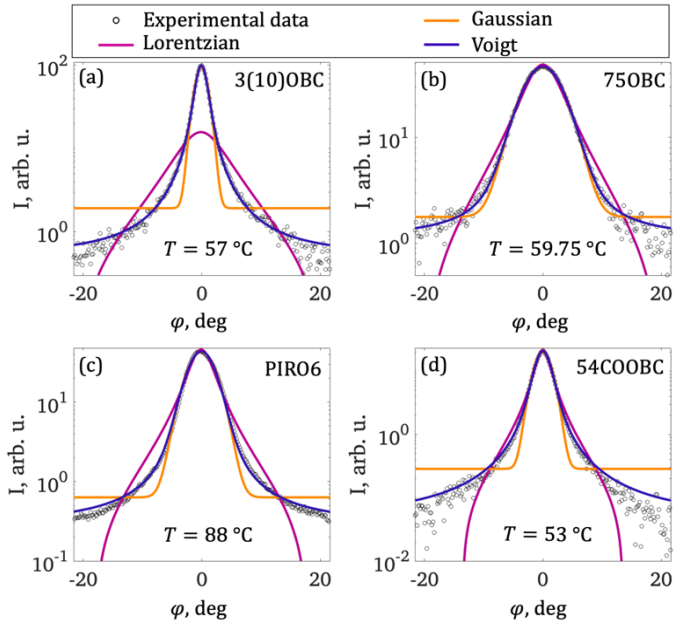

**Fig. 6.** Examples of the fitting of the angular shape of a single hexatic peak in four different LC compounds with the Lorentzian (pink line) $I(\varphi) = B + C \cdot L(\varphi; \kappa)$, Gaussian (yellow line) $I(\varphi) = B + C \cdot G(\varphi; \sigma)$, and Voigt (blue line) $I(\varphi) = B + C \cdot V(\varphi; \sigma, \kappa)$ functions, defined in eqns (5-7). The constant background $B$, the scaling factor $C$, and the angular variables $\sigma$ and $\kappa$ were the free parameters of the fitting.

In Fig. 3 the in-plane positional correlation length $\xi = 1/\gamma$ is shown for four hexatic samples as a function of the relative temperature $\Delta T$, where $\gamma$ is the HWHM of the radial cross section of the scattered intensity determined by the fitting with the Lorentzian function. In the Sm-A phase, $\xi$ is about 1–2 nm and does not change with the temperature. On cooling, upon approaching the Hex-B phase the correlation length $\xi$ starts to increase and shows the maximum of the derivative $d\xi/dT$ at the phase transition point $T_C$[20] (compare with[27]). The temperature dependence of the positional correlation length differs from one hexatic compound to another due to differences in material properties and the values of a coupling constant determining the strength of the interaction between the lateral density fluctuations and the BO order. Thus, the correlation length $\xi$ either saturates deeper in the Hex-B phase upon cooling as in 75OBC and 54COOBC, or linearly increases as in 3(10)OBC and PIRO6 hexatic films.[19,20] Close to the crystallization temperature the in-plane positional correlation length increases by about an order of magnitude. In the crystal phase, the positional order becomes long-range, and the diffraction peaks become so narrow that the analysis of the shape is limited by the resolution of the detector (see for example[18]).

### 3.2 Orientational order and the azimuthal intensity distribution

Now we analyze the shape and the width of individual hexatic scattering peaks in the azimuthal direction, $I(q_{\perp 0}, \varphi) = I(\varphi)$. In Fig. 4, examples of the temperature evolution of the angular profiles for all four Hex-B compounds are shown.[18-20,38] It is readily seen that close to the phase transition point the peaks are relatively broad, which corresponds to the presence of only few first harmonics in the Fourier expansion (2). Upon the temperature decrease, the peaks become more intense and narrower. However, even far away from the Sm-A – Hex-B transition the well-distinguishable Lorentzian-like wings still persist. The origin of these wings lies in the coupling between the positional and BO order (see eqn (4) and the Supplementary Information, section 3). To quantify the temperature evolution of the azimuthal peak profile in various hexatic LCs, we show in Fig. 5 their full width at half maximum (FWHM) as a function of temperature deviation $\Delta T$ from the Sm-A – Hex-B phase transition point. The FWHM of the angular profiles decreases with the temperature by about an order of magnitude, and nearly saturates deep in the Hex-B phase. The temperature-induced changes in the positional and BO order, i.e. abrupt changes close to the phase transition temperature $\Delta T$=0, and a subsequent gradual evolution deeper in the Hex-B phase, occur almost synchronously as it can be seen in Figs. 3 and 5. It further proves the strong coupling between the short-range positional order and long-range BO order.

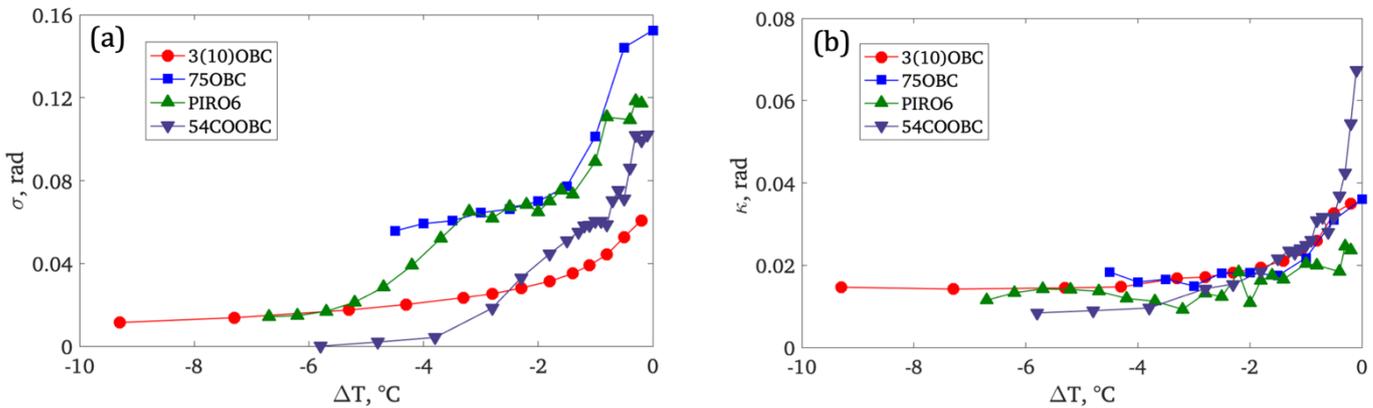

**Fig. 7.** Temperature dependence of the parameters $\sigma$ (a) and $\kappa$ (b) for different LC compounds obtained from the fitting of the azimuthal profiles with the Voigt function $V(\varphi; \sigma, \kappa)$.

In order to get more insight on the temperature dependences of the azimuthal peak shapes in the Hex-B phase $I(\varphi)$, we have approximated these experimentally obtained profiles with three distinct functions, namely the Lorentzian, Gaussian and Voigt functions, defined in eqns (5-7). To fit the experimental data, the normalized functions (5-7) were scaled by a constant factor $C$ and then a constant pedestal $B$ was added. One can see in Fig. 6, that pure Lorentzian and Gaussian functions do not provide good fitting results of the experimental data. On the other hand, in the whole range of temperatures corresponding to the Hex-B phase the Voigt function provides a good fit of the experimental angular profiles, including the wings of the hexatic peak, which is especially clear at low temperatures, when the hexatic peaks become narrow. The reason for this is more substantial than the mere fact that the Voigt function has one more free fitting parameter compared to the Lorentzian or Gaussian functions, as it was discussed in Section 2. The central part of the azimuthal intensity profile clearly has a Gaussian-like shape, as a result of the small fluctuations of the BO order parameter in 3D hexatic films. The contribution from the Lorentzian function has its origin in the short-range positional order and is necessary to describe the long wings of the experimental profiles $I(\varphi)$.

Next, we analyze how the shape of the Voigt function changes with temperature. In Fig. 7 the parameters $\sigma$ and $\kappa$ of the Voigt function are plotted for four different LC compounds. The general trend is clear: the values of $\sigma$ and $\kappa$ monotonically decrease on cooling within the Hex-B phase, while the decay law is different for the two parameters. The value of the parameter $\kappa$ is lower than σ; it shows a steep decrease at small deviations $\Delta T$ from the Sm-A –Hex-B phase transition point, and then stays almost constant at lower temperatures.

To characterize the relative contributions of the Gaussian and Lorentzian functions to the angular profile of the diffraction peak fitted by the Voigt function we have used a dimensionless parameter $\varepsilon = \Delta L/(\Delta L + \Delta G)$, where $\Delta L = 2\kappa$ and $\Delta G = 2\sqrt{2\ln 2}\,\sigma$ are the FWHMs of the Lorentzian and Gaussian functions, respectively.[41] Parameter $\varepsilon$ varies from zero to unity, where $\varepsilon = 0$ corresponds to the pure Gaussian, and $\varepsilon = 1$ corresponds the pure Lorentzian functions.

The temperature dependence of the parameter $\varepsilon$ for four LC hexatic compounds under study is shown in Fig. 8. The behavior of the parameter $\varepsilon$ is different for various LC compounds, which reflects the difference in the coupling constants determining the strength of the interaction between the short-range density fluctuations and the BO order (see Section 2). For the compounds 75OBC and PIRO6, the value of $\varepsilon$ is slightly below 0.2 close to the Hex-B – Sm-A phase transition points and increases up to $\varepsilon \approx 0.4$ deeper in the Hex-B phase (PIRO6). Such a behavior indicates that in the phase transition region the Gaussian contribution to the hexatic angular structure factor dominates, and only upon increasing the temperature deviation $\Delta T$ from the Hex-B – Sm-A phase transition point, the Lorentzian contribution to the Voigt function is approaching the Gaussian counterpart.

Different type of behavior is observed for the 3(10)OBC compound (Fig. 8). In this case, the contributions of the Gaussian and Lorentzian profiles in the total structure factor in the vicinity of the phase transition point is nearly equal (parameter $\varepsilon$ is only slightly below 0.5) and only deeper in the Hex-B phase the Lorentzian profile slowly starts to dominate. The temperature behavior of the 54COOBC compound is quite exceptional: at the Sm-A – Hex-B phase transition region, the Gaussian contribution to the angular structure factor clearly dominates. However, upon further cooling into the Hex-B phase the Lorentzian contribution steeply increases and becomes determining factor in the angular Voigt profile. Such behavior might be related to the particular thermodynamic properties of the 54COOBC compound, characterized by a first-order Sm-A – Hex-B phase transition, which occurs in the close vicinity to the tricritical point between the Sm-A, Hex-B, and crystalline phases.[38]

The general temperature behavior of the parameter $\varepsilon$ shown in Fig. 8 for all four hexatic compounds is in good accordance with the predictions of Aeppli and Bruinsma[21] and the MCST[23] (see Section 2). At the phase transition region, the mean square amplitude of fluctuations of the phase $\langle\delta\psi^2\rangle$ of the BO order parameter is large, and the wide Gaussian function dominates over the Lorentzian function in the convolution integral. Upon cooling down, the temperature deviation $\Delta T$ from the Hex-B – Sm-A phase transition point increases and consequently the magnitude of $\langle\delta\psi^2\rangle$ becomes smaller. Thus, for all LC materials we see an increase of the parameter $\varepsilon$ on cooling (Fig. 8). For 54COOBC the fluctuations of the BO order $\langle\delta\psi^2\rangle$ become so small, that the convolution of a very narrow Gaussian profile with a relatively wide Lorentzian profile results in the Voigt function with a significantly dominating Lorentzian

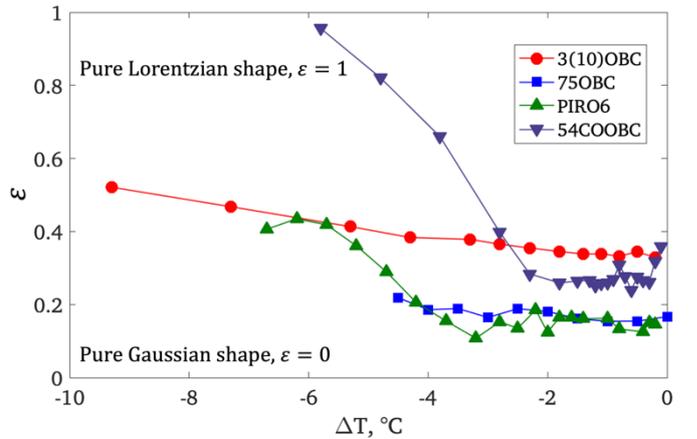

**Fig. 8.** Temperature dependence of the parameter $\varepsilon$ obtained from the fitting of the angular profile of a single hexatic peak with the Voigt function for four hexatic LC compounds (eqns (5-7)). The value $\varepsilon = 0$ corresponds to the pure Gaussian function, and $\varepsilon = 1$ – to the pure Lorentzian function.

contribution.

An interesting observation can be made by comparing the values of parameters $\gamma = 1/\xi$ determined from the radial cross section of the hexatic peak (Fig. 3), and $\kappa$ determined from the azimuthal cross section of the same peak (Fig. 7(b)). Since both parameters are related to the short-range in-plane positional order, one can expect that their ratio is temperature independent. However, the experiment shows that the dimensionless ratio $\gamma/(q_{\perp 0}\kappa)$ actually increases on cooling (see Supplementary Information). Such a behavior can be attributed to the increase of the coupling between the BO order and the short-range positional order on cooling. A more striking feature is that the temperature dependence of the ratio $\gamma/(q_{\perp 0}\kappa)$ seems to follow some universal law for the LC materials with the second-order Sm-A – Hex-B phase transition (3(10)OBC, 75OBC and PIRO6), while for the first-order Sm-A – Hex-B transition (54COOBC) the ratio $\gamma/(q_{\perp 0}\kappa)$ shows much flatter temperature dependence (see Supplementary Information). These trends clearly require additional studies.

## 4 Conclusions

The most important result of our study comes from the analysis of the shape of the angular structure factor in the Hex-B phase for different LC compounds. We analysed the results of the synchrotron X-ray scattering measurements of the free-standing LC films showing the characteristic in-plane pattern with the sixfold reflections from the single hexatic domains. It was found that the angular profiles of the in-plane hexatic scattering are well fitted by the Voigt function that is a convolution of the Gaussian and Lorentzian profiles. The relative contribution to the angular scattering profile from both functions depends on the temperature deviation from the Sm-A – Hex-B transition point. We showed that two theoretical approaches shed light on the origin of the peculiar Hex-B angular profiles. The first is the multicritical scaling theory, MCST, which works remarkably well in pinning together successive harmonics of the BO order and their temperature evolution. We show that the scaling law for the three-dimensional hexatics indeed implies the Voigt function as an angular shape of the scattering peak. At the same time, Aeppli and Bruinsma have developed the Landau theory of the hexatic order in LCs. It was shown that the unique shape of the hexatic angular structure factor is due to the coexistence of two terms in the free energy of the hexatic phase: a liquid-like density term and a mixed term corresponding to the coupling between the BO order and the short-range density fluctuations. Within this approach it can be shown that the angular cross-section of an individual hexatic peak is also a convolution of the Gaussian and Lorentzian functions. On the basis of this theory, we have explicitly shown that the relative contribution of the Gaussian and Lorentzian profiles in the Hex-B angular structure factor is determined by the interplay between the mean square amplitude of fluctuations of the phase $\langle\delta\psi^2\rangle$ of the BO order parameter, as well as parameters characterizing the short-range in-plane positional order. Our experimental X-ray data for various hexatic LC compounds are in good agreement with the above theoretical predictions.

## Author Contributions

I.A.V. and B.I.O. conceived of the presented idea. I.A.Z., R.K., M.S., I.A.V., and B.I.O. performed the X-ray experiments. I.A.Z. and R.K. processed the experimental data, I.A.Z. conducted the analytical calculations. I.A.Z., I.A.V. and B.I.O. wrote the paper. All authors read and agreed on the final text of the paper.

## Conflicts of interest

There are no conflicts to declare.

## Acknowledgements

We acknowledge DESY (Hamburg, Germany), a member of the Helmholtz Association HGF, for the provision of experimental facilities. This research was carried out at PETRA III synchrotron facility and used the Coherence Application P10 beamline. We are thankful to E. I. Kats, V. V. Lebedev, A.R. Muratov and E. S. Pikina for the valuable theoretical discussions. We thank Wim de Jeu, C.C. Huang and E. Gorecka for providing us hexatic substances and for stimulating discussions. The work of I.A.Z. was funded by the Federal Ministry of Education and Research (BMBF) and the Baden-Württemberg Ministry of Science as part of the Excellence Strategy of the German Federal and State Governments. I.A.Z. acknowledges the Zeiss foundation for financial support. The work of B.I.O was supported by the Russian Science Foundation (Grant No. 18-12-00108). The preparation of the LC samples and initial characterization of the Hex-B materials at FSRC "Crystallography and Photonics" RAS were supported by the Ministry of Science and Higher Education of Russian Federation within the corresponding State assignments. We are thankful to E. Weckert and F. Schreiber for support of the project.

# Supplementary information

## 1  Coupling between the bond-orientational order and positional correlations in the hexatic phase

Below we reproduce the arguments presented in the paper by G. Aeppli and R. Bruinsma[1] in order to show that the coupling between the density fluctuations and the bond-orientational (BO) order leads to the angular profile of the scattering peaks in the Hex-B phase described by the Voigt function (see also[2]).

At the temperatures below the Sm-A – Hex-B phase transition, the positional correlations between the molecules in the Hex-B layers decay over a short distance $\xi \sim q_0^{-1}$, while the BO order persists over much larger distances within single hexatic domains, $\Lambda_0^{-1} \gg q_0^{-1}$. Here $q_0$ is a reciprocal vector corresponding to the average distance between the molecules. This allows us to divide the Hex-B into the large cells located at point $r$ with a typical size of $\Lambda_0^{-1}$, in which the two-component BO order parameter $\Psi(r) = |\Psi(r)|e^{i6\psi(r)}$ has approximately constant amplitude $|\Psi(r)|$ and phase $\psi(r)$[1]. We will also assume that the fluctuations of the amplitude $|\Psi|$ of the BO order parameter are relatively small and within the mean-field approximation it can replaced by its root-mean-square value $\langle|\Psi|^2\rangle^{1/2}$. This means that only the phase $\psi(r)$ fluctuates between the cells.

The free-energy functional of the Hex-B phase has three contributions, describing the positional order, the bond-orientational order and the coupling between them:

$$F = F_{\Delta\rho}[\Delta\rho] + F_{\Psi}[\Psi] + F_{\Delta\rho-\Psi}[\Delta\rho, \Psi]. \tag{S1}$$

Let us consider only the in-plane fluctuations of the density, and assume that the molecular layers are perfectly flat. In this case, the first term describing the short-range positional order within a molecular layer can be represented as an integral over the Fourier components of the in-plane density fluctuations $\delta\rho(\boldsymbol{q}_\perp)$

$$F_{\delta\rho}[\delta\rho] = \int d\boldsymbol{q}_\perp \{a + b(q_\perp - q_{\perp 0})^2\}(\delta\rho(\boldsymbol{q}_\perp))^2. \tag{S2}$$

where the integration is performed in the vicinity of the in-plane peak $q_\perp \sim q_{\perp 0}$[3]. Here we used polar coordinates for the scattering vector $\boldsymbol{q} = (q_z, \boldsymbol{q}_\perp) = (q_z, q_\perp, \varphi)$. Coefficients $a$ and $b$ are the coefficients of Landau expansion of the free energy.

Within an individual cell, the two-component BO order parameter $\Psi(r) = |\Psi(r)|e^{i6\psi(r)}$ is approximately a constant, so the contribution from the second term $F_\Psi[\Psi]$ in eqn (S1) is zero. We will take the fluctuations of the phase $\psi(r)$ between the cells into account later (see eqn (S7)).

The third term can be written as a functional of a real-valued function $f$ of three variables[1]

$$F_{\delta\rho-\Psi}[\Delta\rho, \Psi] = \int d\boldsymbol{q}_\perp f\left(q, \langle|\Psi|^2\rangle^{1/2}, \cos[6\varphi - 6\psi(r)]\right)(\delta\rho(\boldsymbol{q}_\perp))^2. \tag{S3}$$

From the Fourier representation of the free energy (eqns (S1-3)), we can find a certain component $\delta\rho(\boldsymbol{q}_\perp)$ of the density fluctuations to the free energy and evaluate the mean squared amplitude of the density fluctuation within a single cell as

$$\langle|\delta\rho(\boldsymbol{q}_\perp)|^2\rangle = \frac{\int d(\delta\rho)\exp\left[-\frac{F_q}{k_B T}\right]|\delta\rho|^2}{\int d(\delta\rho)\exp\left[-\frac{F_q}{k_B T}\right]} = \frac{k_B T}{a + b(q_\perp - q_{\perp 0})^2 + f}. \tag{S4}$$

The same result can be obtained without direct evaluation of the integrals in eqn (S4), but noticing that both contributions to the free energy in eqns (S2) and (S3) are quadratic with respect to the fluctuations $\delta\rho(\boldsymbol{q}_\perp)$, which allows one to use classical equipartition theorem.

To obtain the total structure factor of the hexatic films, the result (S4) has to be averaged over the fluctuations of the BO order parameter $\Psi(r)$, i.e. over the hexatic degree of freedom. In the absence of the coupling term ($f \equiv 0$), the resulting structure factor will be described by a uniform scattering ring with a Lorentzian radial cross section, similar to the Sm-A phase. In the Hex-B phase, $f \neq 0$ and to describe the corresponding structure factor, the small fluctuations of the phase $\psi(r)$ should be taken into account.

To do this it is convenient to use an angular Fourier expansion of $\langle |\delta\rho(q_\perp)|^2 \rangle$ over $(\varphi - \psi(r))$

$$\langle |\delta\rho(q_\perp)|^2 \rangle = \sum_{p=-\infty}^{+\infty} S_{6p} e^{6ip(\varphi-\psi(r))}, \tag{S5}$$

with the Fourier coefficients

$$S_{6p} = \frac{1}{\pi/3} \int_{-\pi/6}^{\pi/6} d\theta\, e^{-6ip\theta} \frac{k_B T}{a + b(q_\perp - q_{\perp 0})^2 + f\left(q, \langle|\Psi|^2\rangle^{1/2}, \cos 6\theta\right)}. \tag{S6}$$

Let us represent the phase $\psi(r)$ of the BO order parameter fluctuating around the mean value $\psi_0$ as $\psi(r) = \psi_0 + \delta\psi(r)$, and perform averaging over the fluctuations $\delta\psi(r)$ independently for each term in the angular Fourier expansion

$$S(q_\perp) = \sum_{p=-\infty}^{+\infty} S_{6p} e^{6ip(\varphi-\psi_0)} \langle e^{-6ip\delta\psi(r)} \rangle. \tag{S7}$$

We will consider the case of low temperature hexatic phase, when the fluctuations are small, $\delta\psi(r) \approx 0$. Thus, we can use the Taylor expansion and keep only the first non-zero term

$$\langle e^{-6ip\delta\psi(r)} \rangle \approx \langle 1 - 6pi\delta\psi(r) - \frac{(6p\delta\psi(r))^2}{2} \rangle = 1 - 18p^2 \langle \delta\psi^2 \rangle \approx e^{-18p^2 \langle \delta\psi^2 \rangle}. \tag{S8}$$

Here we assumed that the mean value $\langle \delta\psi(r) \rangle = 0$ because the fluctuations $\delta\psi(r)$ and $-\delta\psi(r)$ have the same contribution to the free-energy and therefore are equally probable. The root-mean-square value $\langle \delta\psi^2 \rangle = \langle (\psi - \psi_0)^2 \rangle$ can be estimated in the frames of the $x - y$ model in 2D and 3D[4,5].

Combining eqns (S6) and (S7), the structure factor can be rearranged as

$$S(q_\perp) = \sum_{p=-\infty}^{+\infty} S_{6p} e^{6ip(\varphi-\psi_0)} e^{-18p^2\langle\delta\psi^2\rangle}$$

$$= \frac{3}{\pi} \int_{-\pi/6}^{\pi/6} d\theta \left[ \frac{k_B T}{a + b(q_\perp - q_{\perp 0})^2 + f\left(q, \langle|\Psi|^2\rangle^{1/2}, \cos 6\theta\right)} \sum_{p=-\infty}^{+\infty} e^{-6ip(\varphi-\psi_0-\theta)} e^{-18p^2\langle\delta\psi^2\rangle} \right]. \tag{S9}$$

It is easy to check that the sum under the integral converges to a Gaussian function (compare with eqns (S22-S27) in the next section):

$$\frac{3}{\pi} \sum_{p=-\infty}^{+\infty} e^{-6ip(\varphi-\psi_0-\theta)} e^{-18p^2\langle\delta\psi^2\rangle} = \frac{1}{\sqrt{2\pi\langle\delta\psi^2\rangle}} \sum_{n=-\infty}^{\infty} \exp\left[ -\frac{\left(\theta - (\varphi-\psi_0) - \frac{\pi}{3}n\right)^2}{2\langle\delta\psi^2\rangle} \right]. \tag{S10}$$

Since integration over the angular variable $\theta$ in eqn (S9) is performed over a limited range from $-\frac{\pi}{6}$ to $\frac{\pi}{6}$, only one term in the sum over $n$ will contribute to the hexatic structure factor (because $\langle\delta\psi^2\rangle$ is small and there is no overlap between various terms). This can be taken into account by selecting the reference axis in such a way that $\psi_0 = 0$ and considering the scattering in the same direction, i.e. $|\varphi| < \pi/6$. In this case, we can rewrite the hexatic structure factor as

$$S(\boldsymbol{q}_\perp) = \frac{k_B T}{\sqrt{2\pi\langle\delta\psi^2\rangle}} \int_{-\pi/6}^{\pi/6} d\theta \frac{e^{-\frac{(\theta-\varphi)^2}{2\langle\delta\psi^2\rangle}}}{a + b(q_\perp - q_{\perp 0})^2 + f\left(q, \langle|\Psi|^2\rangle^{1/2}, \cos 6\theta\right)}. \quad (S11)$$

Due to the sharp Gaussian peak in the nominator, the main contribution to $S(\boldsymbol{q}_\perp)$ comes from the region around $\theta \approx 0$. This allows us to expand the coupling function $f$ into the Taylor series

$$f\left(q, \langle|\Psi|^2\rangle^{1/2}, \cos 6\theta\right) \approx f_0 + f_2 \theta^2 \quad (S12)$$

The integration region in (S11) can be formally extended to infinity, because contribution from the large angles ($|\theta| > \pi/6 \gg \sqrt{\langle\delta\psi^2\rangle}$) is negligibly small:

$$S(\boldsymbol{q}_\perp) = \frac{k_B T}{\sqrt{2\pi\langle\delta\psi^2\rangle}} \int_{-\infty}^{+\infty} d\theta \frac{e^{-\frac{(\theta-\varphi)^2}{2\langle\delta\psi^2\rangle}}}{a + f_0 + b(q_\perp - q_{\perp 0})^2 + f_2 \theta^2}. \quad (S13)$$

If $\langle\delta\psi^2\rangle \ll (a + f_0)/f_2$, one can expand the denominator into the Taylor series and keep only the first term

$$\frac{1}{a + f_0 + b(q_\perp - q_{\perp 0})^2 + f_2 \theta^2} \approx \frac{1}{a + f_0 + b(q_\perp - q_{\perp 0})^2}\left[1 - \frac{f_2 \theta^2}{a + f_0 + b(q_\perp - q_{\perp 0})^2}\right] \quad (S14)$$

After this simplification, the integral (S13) can be easily evaluated, the radial cross section ($\varphi = 0$) of the diffraction peak from the hexatic phase can be described by a Lorentzian function

$$S(q_\perp) = \frac{k_B T}{a + f_0 + b(q_\perp - q_{\perp 0})^2}\left[1 - \frac{f_2 \langle\delta\psi^2\rangle}{a + f_0 + b(q_\perp - q_{\perp 0})^2}\right] \approx \frac{k_B T}{a + f_0 + f_2\langle\delta\psi^2\rangle + b(q_\perp - q_{\perp 0})^2}$$
$$\propto \frac{1}{\gamma^2 + (q_\perp - q_{\perp 0})^2} \quad (S15)$$

with the half width at half maximum $\gamma = \sqrt{(a + f_0 + f_2\langle\delta\psi^2\rangle)/b}$.

The azimuthal profile through the maximum of the diffraction peak ($q_\perp = q_{\perp 0}$) is given by the Voigt function which is a convolution of the Gaussian and Lorentzian functions

$$S(\varphi) = \frac{k_B T}{\sqrt{2\pi\langle\delta\psi^2\rangle}} \int_{-\infty}^{+\infty} d\theta \frac{e^{-\frac{(\theta-\varphi)^2}{2\langle\delta\psi^2\rangle}}}{a + f_0 + f_2 \theta^2} = \pi \frac{k_B T}{\sqrt{f_2(a + f_0)}} G(\varphi; \sigma) * L(\varphi; \kappa) = \frac{k_B T \pi}{\sqrt{f_2(a + f_0)}} V(\varphi; \sigma, \kappa). \quad (S16)$$

Here the Gaussian function

$$G(\varphi; \sigma) = \frac{1}{\sqrt{2\pi\sigma^2}} e^{-\frac{\varphi^2}{2\sigma^2}} \quad (S17)$$

is defined by the parameter $\sigma = \sqrt{\langle\delta\psi^2\rangle}$ which depends only on the mean squared fluctuations of the phase $\psi(\boldsymbol{r})$. The Lorentzian function

$$L(\varphi; \kappa) = \frac{\kappa}{\pi(\varphi^2 + \kappa^2)} \quad (S18)$$

has the half width at half maximum $\kappa = \sqrt{(a + f_0)/f_2}$ which has contribution from the short-range position order (coefficient $a$ in the Landau expansion of the free energy (2)) and the coupling between the positional and BO order (coefficients $f_0$ and $f_2$ in eqn (S12)).

## 2    Angular profile of the hexatic peak in the multicritical scaling theory

A natural way to describe the BO order in the hexatic phase is to expand the azimuthal dependence of the structure factor into the Fourier series[6]

$$I(q, \varphi) = I_0(q) \left[1 + 2 \sum_{m=1}^{\infty} C_{6m}(q) \cos(6m(\varphi - \varphi_0))\right], \quad (S19)$$

where the coefficients $0 \leq C_{6m} \leq 1$ depend on the degree of the orientational order. The multicritical scaling theory predicts the following relation between the coefficients $C_{6m}$ of different order

$$C_{6m} = (C_6)^{m+\lambda m(m-1)}, \quad (S20)$$

where $\lambda \approx 0.3$ for 3D hexatic phase,[6,7] and $\lambda \approx 1$ for 2D hexatic phase.[4]

First, let us show that the series (S19) converges to a Gaussian function in the 2D case ($\lambda = 1$). Indeed,

$$1 + 2 \sum_{m=1}^{\infty} C_{6m} \cos(6m(\varphi - \varphi_0)) = 1 + 2 \sum_{m=1}^{\infty} (C_6)^{m^2} \cos(6m(\varphi - \varphi_0)) = $$

$$\sum_{m=-\infty}^{\infty} (C_6)^{m^2} e^{i6m(\varphi - \varphi_0)} = \sum_{m=-\infty}^{\infty} \exp\left[-m^2 \ln\frac{1}{C_6} + i \cdot 6m(\varphi - \varphi_0)\right]. \quad (S21)$$

At the same time, the azimuthal profile of the six hexatic peaks, each described by a Gaussian function, can be written as

$$I_G(\varphi) = \frac{I_0}{\sqrt{2\pi\sigma^2}} \sum_{n=-\infty}^{\infty} \exp\left[-\frac{\left(\varphi - \varphi_0 - \frac{2\pi}{6}n\right)^2}{2\sigma^2}\right], \quad (S22)$$

where $\varphi_0$ defines the angular position of the peaks with respect to some reference axis, and the term $\frac{2\pi}{6}n$ appears due to the sixfold symmetry of the hexatic structure factor. The periodic function (S22) can be expanded into the Fourie series

$$I_G(\varphi) = \sum_{m=-\infty}^{\infty} C_{6m} e^{i6m\varphi} \quad (S23)$$

with the Fourier coefficients

$$C_{6m} = \frac{1}{2\pi/6} \int_{\varphi_0 - \frac{\pi}{6}}^{\varphi_0 + \frac{\pi}{6}} I_G(\varphi) e^{-i6m\varphi} d\varphi = \frac{3I_0}{\pi\sqrt{2\pi\sigma^2}} \sum_{n=-\infty}^{\infty} \int_{\varphi_0 - \frac{\pi}{6}}^{\varphi_0 + \frac{\pi}{6}} \exp\left[-\frac{\left(\varphi - \varphi_0 - \frac{2\pi}{6}n\right)^2}{2\sigma^2} - i6m\varphi\right] d\varphi. \quad (S24)$$

Assuming that the hexatic peaks are sharp and do not overlap, we can neglect all terms with $n \neq 0$ (they correspond to the peaks of $I_G(\varphi)$ outside of the integration region). Then we can extend the region of integration to infinity

$$C_{6m} = \frac{3I_0}{\pi\sqrt{2\pi\sigma^2}} \int_{-\infty}^{+\infty} \exp\left[-\frac{(\varphi - \varphi_0)^2}{2\sigma^2} - i6m\varphi\right] d\varphi, \quad (S25)$$

since the main contribution to the integral (S24) comes from the vicinity of the peak at $\varphi = \varphi_0$.

Now substituting $\varphi = t \cdot \sigma\sqrt{2} + \varphi_0$, we can evaluate the coefficients

$$C_{6m} = \frac{3I_0\sigma\sqrt{2}}{\pi\sqrt{2\pi\sigma^2}} e^{-i6m\varphi_0} e^{-18\sigma^2 m^2} \int_{-\infty}^{+\infty} e^{-(t+i3m\sigma\sqrt{2})^2} dt = \frac{3I_0}{\pi} e^{-i6m\varphi_0} e^{-18\sigma^2 m^2}. \tag{S26}$$

Thus, the Fourier series (S23) can be written as

$$I_G(\varphi) = \frac{3I_0}{\pi} \sum_{m=-\infty}^{\infty} \exp[-18m^2\sigma^2 + i \cdot 6m(\varphi - \varphi_0)], \tag{S27}$$

which coincides with the multicritical theory expansion (S21) up to the prefactor, assuming $C_6 = e^{-18\sigma^2}$. Since the Fourier expansion is unique, we can conclude that the scaling law (S20) with $\lambda = 1$ (2D case) corresponds to the Gaussian shape of the hexatic peaks in the azimuthal direction.

In the 3D case ($\lambda \approx 0.3$), the angular Fourier series in eqn (S19) can be written as

$$1 + 2\sum_{m=1}^{\infty} C_{6m} \cos(6m(\varphi - \varphi_0)) = \sum_{m=-\infty}^{\infty} (C_6)^{\lambda m^2 + (1-\lambda)|m|} e^{i6m(\varphi-\varphi_0)} =$$

$$\sum_{m=-\infty}^{\infty} \exp\left[-m^2\lambda \ln\frac{1}{C_6} - |m|(1-\lambda)\ln\frac{1}{C_6} + i \cdot 6m(\varphi - \varphi_0)\right] = \tag{S28}$$

$$\sum_{m=-\infty}^{\infty} \exp[-18m^2\sigma^2 - 6\kappa|m| + i \cdot 6m(\varphi - \varphi_0)]$$

where the parameters $\sigma$ and $\kappa$ can be expressed through the parameters of the multicritical scaling theory:

$$\sigma = \sqrt{\frac{\lambda}{18} \ln\frac{1}{C_6}},$$
$$\kappa = \frac{1-\lambda}{6} \ln\frac{1}{C_6}. \tag{S29}$$

Now the angular Fourier coefficients $C_{6m}$ are a product of $\exp[-m^2\sigma^2]$, which corresponds to the Fourier coefficients of a Gaussian function (see the derivation above), and $\exp[-6\kappa|m|]$, which corresponds to the Fourier coefficients of a Lorentzian function. Let us prove this by expanding the periodic Lorentzian function into the Fourier series

$$I_L(\varphi) = I_0 \frac{\kappa}{\pi} \sum_{n=-\infty}^{\infty} \frac{1}{\left(\varphi - \varphi_0 - \frac{2\pi}{6}n\right)^2 + \kappa^2} = \sum_{m=-\infty}^{\infty} C_{6m} e^{i6m\varphi}, \tag{S30}$$

and evaluating the Fourier coefficients in a similar way, as it was done for the Gaussian function in eqns (S24-S26):

$$C_{6m} = \frac{1}{2\pi/6} \int_{\varphi_0-\frac{\pi}{6}}^{\varphi_0+\frac{\pi}{6}} I_L(\varphi) e^{-i6m\varphi} d\varphi = \frac{3\kappa I_0}{\pi^2} \sum_{n=-\infty}^{\infty} \int_{\varphi_0-\frac{\pi}{6}}^{\varphi_0+\frac{\pi}{6}} \frac{e^{-i6m\varphi}}{\left(\varphi - \varphi_0 - \frac{2\pi}{6}n\right)^2 + \kappa^2} d\varphi = \frac{3\kappa I_0}{\pi^2} \int_{-\infty}^{+\infty} \frac{e^{-i6m\varphi}}{(\varphi - \varphi_0)^2 + \kappa^2} d\varphi$$

$$= \frac{3\kappa I_0}{\pi^2} e^{-i6m\varphi_0} \int_{-\infty}^{+\infty} \frac{e^{-i6m\varphi}}{\varphi^2 + \kappa^2} d\varphi = \frac{3\kappa I_0}{\pi^2} e^{-i6m\varphi_0} \frac{\pi}{\kappa} e^{-\kappa|6m|} = \frac{3I_0}{\pi} e^{-6\kappa|m| - i6m\varphi_0}. \tag{S31}$$

Therefore, the Fourier expansion of the Lorentzian function (S30) can be written as

$$I_L(\varphi) = \frac{3I_0}{\pi} \sum_{m=-\infty}^{\infty} \exp[-6\kappa|m| + i \cdot 6m(\varphi - \varphi_0)], \tag{S32}$$

proving that the term $\exp[-6\kappa|m|]$ in eqn (S28) indeed corresponds to the Lorentzian function.

Using the convolution theorem, we can expect that the Fourier series (S28), in which each term is equal to the product of the Fourier components of the Gaussian and Lorentzian functions, will converge to a convolution of these two functions.

Therefore, from eqns (S19) and (S28) it follows that for the 3D hexatics the angular distribution of intensity can be written as

$$I_V(\varphi) = I_0 \sum_{n=-\infty}^{\infty} V\left(\varphi - \varphi_0 - \frac{2\pi}{6}n; \sigma, \gamma\right), \qquad (S33)$$

where $V(\varphi; \sigma, \kappa)$ is the Voigt function

$$V(\varphi; \sigma, \gamma) = G(\varphi; \sigma) * L(\varphi; \kappa) = \int_{-\infty}^{+\infty} d\theta \frac{e^{-\frac{\theta^2}{2\sigma^2}}}{\sqrt{2\pi\sigma^2}} \cdot \frac{\frac{\kappa}{\pi}}{(\varphi - \theta)^2 + \kappa^2}, \qquad (S34)$$

which is a convolution of a Gaussian $G(\varphi; \sigma)$ and a Lorentzian $L(\varphi; \kappa)$ functions defined in eqns (S17-S18).

## 3  Correlations between two short-range order parameters $\gamma$ and $\kappa$

The short-range positional order in the structure factor of the Hex-B phase is described by two different parameters: $\gamma$ and $\kappa$. The first one, $\gamma = 1/\xi$, defines the width of the Lorentzian profile in the radial cross section through the peak of the structure factor, while the second one, $\kappa$, defines the width of the Lorentzian contribution to the Voigt profile in the azimuthal direction for the same peak (eqns S16-S18). The common physical origin of these two parameters arises a natural question about the possible correlation between them. In order to analyze this question, we used the experimental data from the Fig 3 and Fig 7(b) of the main text to determine a dimensionless ratio $\gamma/(\kappa q_{\perp 0})$, shown in Fig S1.

This ratio is zero at the phase transition point ($\Delta T = 0$ °C), because the coupling between the BO order and the positional order is absent in the Sm-A phase, so there is no angular modulation of the scattered intensity, i.e. $\kappa$ is infinite. For 54COOBC compound, the ratio $\gamma/(\kappa q_{\perp 0})$ reaches the value of 0.3-0.4 on cooling and stays almost constant. Such a behavior strongly supports our hypothesis that both parameters $\gamma$ and $\kappa$ are strongly correlated, i.e. their ratio is constant.

However, for three other LC materials, the situation looks different and the saturation is achieved at much lower temperatures ($\Delta T \leq -2$ °C) and for the higher values of the ratio ($\gamma/(\kappa q_{\perp 0}) = 3 - 4$). This behavior can be explained by the fact that the parameters $\gamma$ and $\kappa$ depend on the Landau coefficients $a$, $b$, $f_0$, $f_2$ as well as the mean square fluctuations of the phase of the BO order $\langle \delta\psi^2 \rangle$ (see eqns (S15-S18)). The value of all these parameters may depend on temperature, therefore the accurate analysis of the temperature dependence of the ratio $\gamma/(\kappa q_{\perp 0})$ is a complex problem, which requires a dedicated study. Here we would like to point out the striking similarity between the temperature dependence of the ratio $\gamma/(\kappa q_{\perp 0})$ for three different LC compounds (3(10)OBC, 75OBC and PIRO6) showing a second-order Sm-A – Hex-B phase transition. This might be an indication of some universal law governing the coupling between the positional and the BO order in the vicinity of the Sm-A – Hex-B transition of the second order. Such a universal behavior is clearly absent for the case of 54COOBC compound exhibiting the first-order Sm-A – Hex-B phase transition.

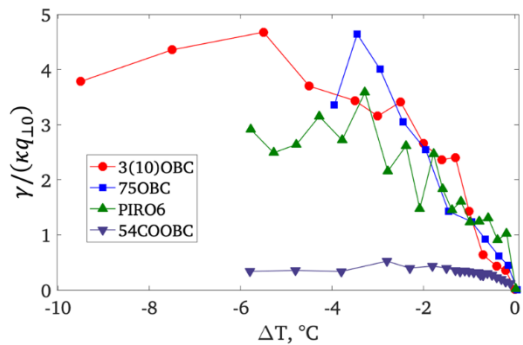

Fig. S1. The dimensionless ratio $\gamma/(q_{\perp 0}\kappa)$ as a function of the temperature $\Delta T = T - T_C$ measured relative to the phase transition temperature $T_C$.